# From leaf to tree: upscaling of artificial photosynthesis

*Bugra Turan\*, Jan-Philipp Becker, Félix Urbain, Friedhelm Finger, Uwe Rau, and Stefan Haas*

IEK5 - Photovoltaik, Forschungszentrum Jülich GmbH, 52425 Jülich, Germany
E-mail: b.turan@fz-juelich.de

**Abstract**

Energy storage becomes crucial for energy systems with an increasing share of renewable energy sources. Artificial photosynthesis, in particular photovoltaic water splitting, provides both sustainable energy generation and energy storage in the form of hydrogen. However, only a few concepts for scalable devices were reported in the literature. Here, we introduce a new concept which, by design, is scalable and compatible with every thin-film photovoltaic technology. The concept allows for independent geometrical optimization of the photovoltaic and the electrochemical part. The scalability is achieved by continuous mirroring of a base unit. We demonstrate a fully integrated, wireless device with a stable and bias-free operation for 40 hours. The concept was scaled to an area of 64 cm² comprising 13 base units and exhibited a solar-to-hydrogen efficiency of 3.9%. The concept and its successful realization is an important contribution towards the large scale application of artificial photosynthesis.

# Introduction

In the last decades, renewable energy sources gained a significant share in the world-wide energy supply. Due to the fluctuating nature of renewable energy sources the challenges for further implementation gradually shift from energy generation to energy storage. This trend becomes especially obvious in the fact that research on direct photo-catalytic water splitting, which as a research topic dates back to the 1970s[1], has regained considerable interest during recent years[2–6]. This research progress is driven not only by advancements in materials science concerning new photovoltaic absorber materials[7] and novel catalysts[8–11], but also by design optimizations of multiple junction solar cells[12–15].

However, in view of the urgent need to develop the technology towards large scale applications, it appears staggering that research still is almost exclusively focused on laboratory experiments. Comparatively little effort has been devoted to the design and realization of large area or at least scalable devices[16,17]. Taking the recent progress of photovoltaic energy conversion as a paradigm for successful implementation of a renewable energy technology[18–20], it becomes clear that in the end cost effectiveness becomes as much a question of clever device design and process engineering as a question of optimized components. This especially applies for water splitting devices which require both a proper management of photons and electrons in the photovoltaic part and of the ions in the electrochemical part. Moreover, many trade-offs between optimizing different components show up only in view of completed, fully integrated devices.

The present paper introduces the design and realization of monolithically integrated solar water splitting modules based on silicon thin-film module technology. The realized devices fulfill the basic requirements for a future large scale technology, i.e. they are wireless and perfectly scalable to arbitrary device areas. The scalability is achieved by a continuous repetition of a base unit which in itself combines a photovoltaic (sub) device with the two electrodes of an electrolyzer.

We took advantage of the laser patterning processes used for the series connection of thin-film solar cells[21–23] and the wide range of design options going along with this type of processing. Three devices were realized. The base unit of the photovoltaic water splitting device utilized either a series connection of three a-Si:H single junction cells or two a-Si:H/μc-Si:H tandem cells connected in series. The first device was a singular base unit designed to access the photovoltaic and electrochemical data individually and to evaluate the faradaic efficiency. The second device was built to investigate the operation stability. The third device consisted of a large area module (device area $A = 64$ cm²) encompassing 13 base units. The latter represents one of the very few practical demonstrations of a scalable monolithic water splitting concept reported in the literature.

## General design considerations

Any form of solar photo-electrochemical water splitting device comprises a series connection of a photovoltaic cell (PV) that converts solar photons in electrons and holes,

$$h\nu \rightarrow e^- + h^+, \tag{1}$$

and an electrochemical cell (EC) with two half-reactions. In an alkaline electrolyte the hydrogen evolution reaction (HER) at the cathode is given by

$$2\ H_2O + 2e^- \rightarrow H_2 + 2\ OH^- \tag{2}$$

while the oxygen evolution reaction (OER) reaction at the anode is given by

$$4\ OH^- + 4h^+ \rightarrow O_2 + 2\ H_2O. \tag{3}$$

The electrical circuit is closed via transport of OH⁻ ions in the electrolyte, whereas $H_2$ and $O_2$ are evolving at the cathode and the anode, respectively. Finally, a membrane or a glass frit can prevent mixing of the two gases.

Water splitting only takes place if the solar generated electron and holes have enough energy to overcome the energetic barrier of water oxidation, the overpotential losses at anode and cathode and ohmic losses in the device. For state-of-the-art catalysts, a voltage of approx. 1.7 V to 1.8 V is needed for a current density of 10 mA/cm² at the electrodes[8]. Such high voltages can be generated by single junction solar cells with high band gap energy[24] (approx. $E_g \gtrsim 2.2$ eV) or by the use of multiple junctions connected in series[14]. Since the optimal band gap for single junction solar cells under AM1.5G illumination is 1.34 eV the use of high band gap solar cells leads to high optical transmission losses and thus to inefficient solar cells[25,26]. In contrast, the utilization of multi-junction solar cells, spatially neighboring interconnected solar cells, or a combination of both offers more freedom with respect to the optimization of the device[27]. In the present work, both types of series connection approaches were used.

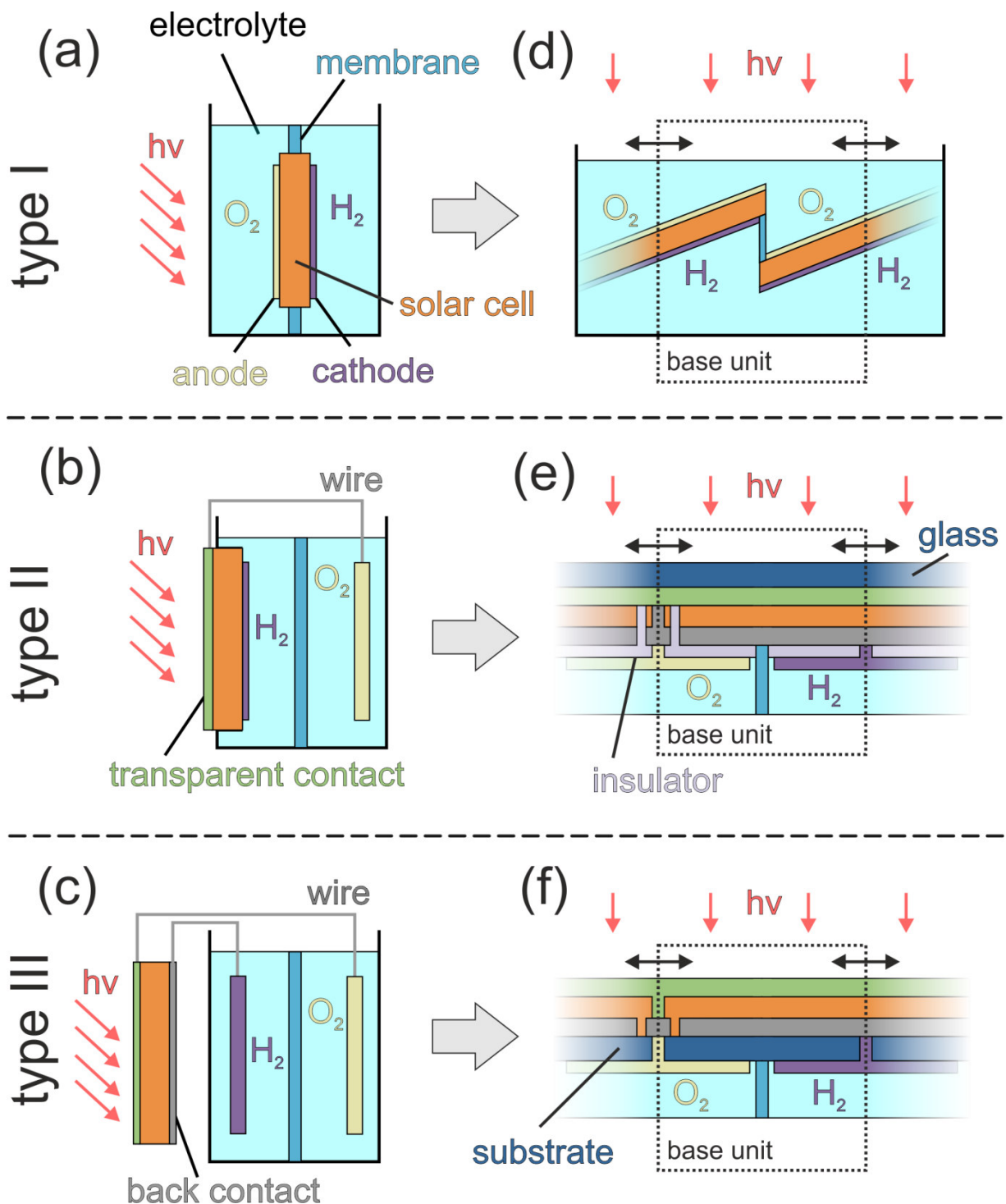

**Figure 1.** Classification of different photovoltaic water splitting designs typically used in laboratory experiments (a)-(c) and the corresponding device designs of possible large devices (d)-(f). For the laboratory experiment (a) and the upscaled design (d) the photovoltaic (PV) part is completely immersed in the electrolyte with the catalyst electrodes on both sides (type I). For type II, (b) and (e), only one side of the PV part is exposed to the electrolyte. Note that a wire connection between the PV front contact of the PV element and the back electrode of the electrochemical cell (EC) is needed for the laboratory experiment (b), whereas the monolithic design of the scalable device is wireless (e). Type III represents an approach with completely separated PV and EC elements. In the laboratory setup (c) two wire connections are needed. The corresponding monolithic (f) approach needs electrical contacts through the substrate but is, again, wireless.

As shown in Figure 1, all laboratory experiments may be categorized into three types, dependent on the number of wires required. Additionally, Figure 1 illustrates how all three types can be converted into a scalable module design. The upscaling is achieved by a continuous repetition and/or mirroring of one base unit as indicated by the dashed boxes in Figure 1(d)-(f).

Figure 1(a) depicts an approach where the whole device is immersed in the electrolyte (type I). Examples for this design were reported by Lin et al.[28] and by Reece et al.[29]. A scalable adaption via the 'louvered cell' (cf. Figure 1(d)) design has been introduced by Walczak et al.[17] using Si and $WO_3$ as photoabsorbers. This device represents one of the rare practical examples of a water splitting module consisting of more than one base unit (two units in the demonstrated design). So far a solar-to-hydrogen (STH) efficiency of only $\eta_{STH}$ = 0.24% has been demonstrated for an operation time of more than 24 hours.

Figure 1(b) shows a design with one wire (type II). This approach is often used for systems embracing superstrate thin-film solar cells (illumination through the glass support)[30–32]. This has the advantage that optical and electrochemical properties can be optimized independently due to the spatial separation. Solar cells based on superstrate technology can be transformed to a scalable water splitting device by the approach depicted in Figure 1(e) with a coplanar electrode configuration. A realization of a base unit employing a-Si:H/a-Si:H tandem cells has been demonstrated as early as 1985 by Appleby et al.[33]. They reported a solar-to-hydrogen efficiency of $\eta_{STH}$ = 2.6%. The series connection of the two tandem cells in the unit was achieved by wiring. Yamada et al. reported a wireless base unit using a thin-film silicon triple junction solar cell[34].

Finally, it is possible to keep the photovoltaic unit completely outside of the electrolyte (Figure 1(c), type III) using two wires. On the laboratory scale, this approach is used to illustrate the compatibility of solar cell technologies with an electrochemical cell of the same area[35–37]. As shown in Figure 1(f) an upscaling of this configuration is also feasible. Recently, Jacobsson et al. have demonstrated an intermediate step to a scalable substrate device based on the series connection of three $Cu(In,Ga)Se_2$ solar cells which exhibits an efficiency of $\eta_{STH}$ = 10 %[38]. However, in this device metal stripes were used for the electrical connection to the water splitting electrodes. For an arbitrarily scalable device, the electrical connections could be realized by holes through the substrate as illustrated in Figure 1(f).

Each of the introduced upscaling concepts exhibits inherent advantages and disadvantages. Apparent disadvantages of type I (Figure 1(d)) are the inevitable illumination through the electrolyte[39] and the complex mechanical construction. These drawbacks are avoided by type II and III. However, for type II an additional electrical insulation is required and for type III contact formation through the substrate can be challenging.

For the practical demonstration of a scalable, wireless, and monolithic solar water splitting device we chose type II because this concept can be realized using mature technologies for large scale production. Furthermore, scarce materials were avoided.

## Device realization

For the present work we realized solar water splitting modules following the design concept shown in Figure 1(e). The devices were either based on three series connected single junction a-Si:H (p-i-n) solar cells or two series connected a-Si:H/µc-Si:H (p-i-n/p-i-n ) tandem cells to generate the required voltage to sustain electrolysis. We chose these different types of absorber configurations to illustrate the flexibility of the design. However, the design can also be used with high band gap absorbers or stacked solar cells with higher voltages to generate the needed voltage without a lateral side-by-side series connection.

Figure 2 shows a cross sectional sketch of the proposed concept for the case of three spatially neighboring series connected solar cells used as a base unit. The PV element was designed in superstrate configuration which is often used in thin-film silicon, CdTe[40], dye-sensitized[41], or organo-metal halide perovskite[42] solar cells.

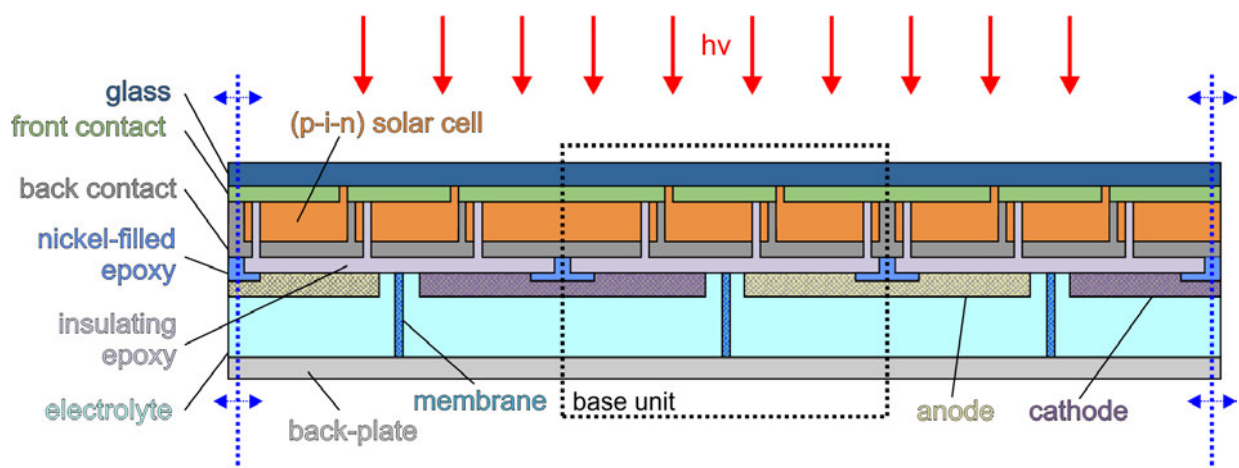

**Figure 2. Cross section of a scalable, fully integrated photovoltaic water splitting device in superstrate configuration. The number of cell stripes in series can be easily adjusted (three in this case). Dimensions are not to scale in width and thickness of the layers. The sketch only shows an excerpt of the module. The configuration can be extended in both directions (hinted by the dashed blue arrows). The base unit that defines the region of periodic repetition/mirroring is depicted by the dashed box.**

The series connection of the individual solar cells was realized by the introduction of a structure that connects the back contact of one cell with the front contact of the adjacent cell. This structure was created by selective laser ablation in between the layer deposition steps.

As apparent in Figure 2, the anodes and cathodes were placed side-by-side on the back side of the PV element. Note that neighboring base units share electrodes. Hence, the sequence of the interconnection needed to be alternating. Thereby, the design avoids additional active area losses due to further laser scribes.

The insulating epoxy layer was used as a corrosion protection against to the alkaline electrolyte as well as for electrical insulation. Chemically resistant epoxy resins have proven to be stable when 1M KOH was used as electrolyte. Additionally, a conductive polymer resin was applied at the electrical interfaces between PV element and the EC anodes/cathodes. Nickel-filled epoxy fulfilled the requirements for both chemical protection and electrical conductivity.

For the sake of simplicity, bare nickel-foam was used for both anode and cathode. Furthermore, for this proof of concept no membrane was employed.

The PV front end and the EC back end can be optimized independently because they are merely coupled by the insulating and conductive polymer coatings. Thus, the concept allows for the utilization of various PV and EC technologies.

**Faradaic efficiency (device #1)**

In a first step the presented concept was realized by processing of one base unit (refer to as device #1) with a design similar to the one sketched in Figure 2 (dashed box). A module with three a-Si:H solar cells connected in series was used here.

For the evaluation of the faradaic efficiency the correlation of the gas generation rate to the current through the system is required. However, the wireless design of the device merely allows monitoring of the operating voltage point of the system. Therefore, a decoupling of the PV and EC elements was necessary. This was achieved by the use of an insulating epoxy layer instead of the nickel-filled epoxy depicted in Figure 2. Wires were soldered to the contacts of the PV and EC elements. Thereby, the current can be monitored as well. Figure 3 shows a photograph of such a module and a block diagram illustrating the different measurement modes.

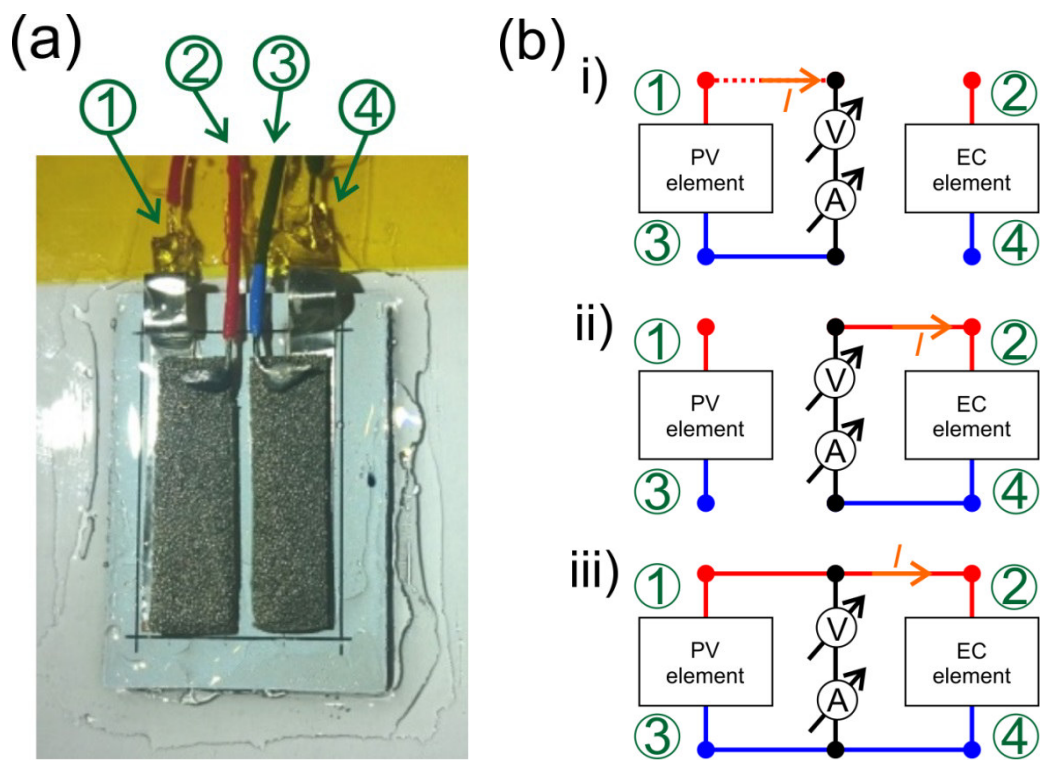

**Figure 3** (a) Photograph of device #1 with separated, electrically insulated, contacts for characterization of the individual properties of PV and EC element. A series connection of three cell stripes, each with a length of 27 mm and a width of 5 mm lead to a device area of 4.8 cm$^2$. (b) Block diagram of the device and different operating modes i) for measuring the photovoltaic part, ii) for measuring the electrochemical part, and iii) for the combined measurement. The numbers in green depict the device terminals.

As shown in Figure 3(b) three operating modes of the device are possible: i) PV operation, ii) EC operation, and iii) simultaneous monitoring of the current and voltage during water splitting.

Eventually, this device was used to determine the faradaic efficiency of the system by a correlation between the experimental gas rate and the theoretical gas rate calculated from the current during operation. The graphs in Figure 4(a) depict the current-voltage (*I-V*) characteristics of this monitoring module after 3 hours of operation.

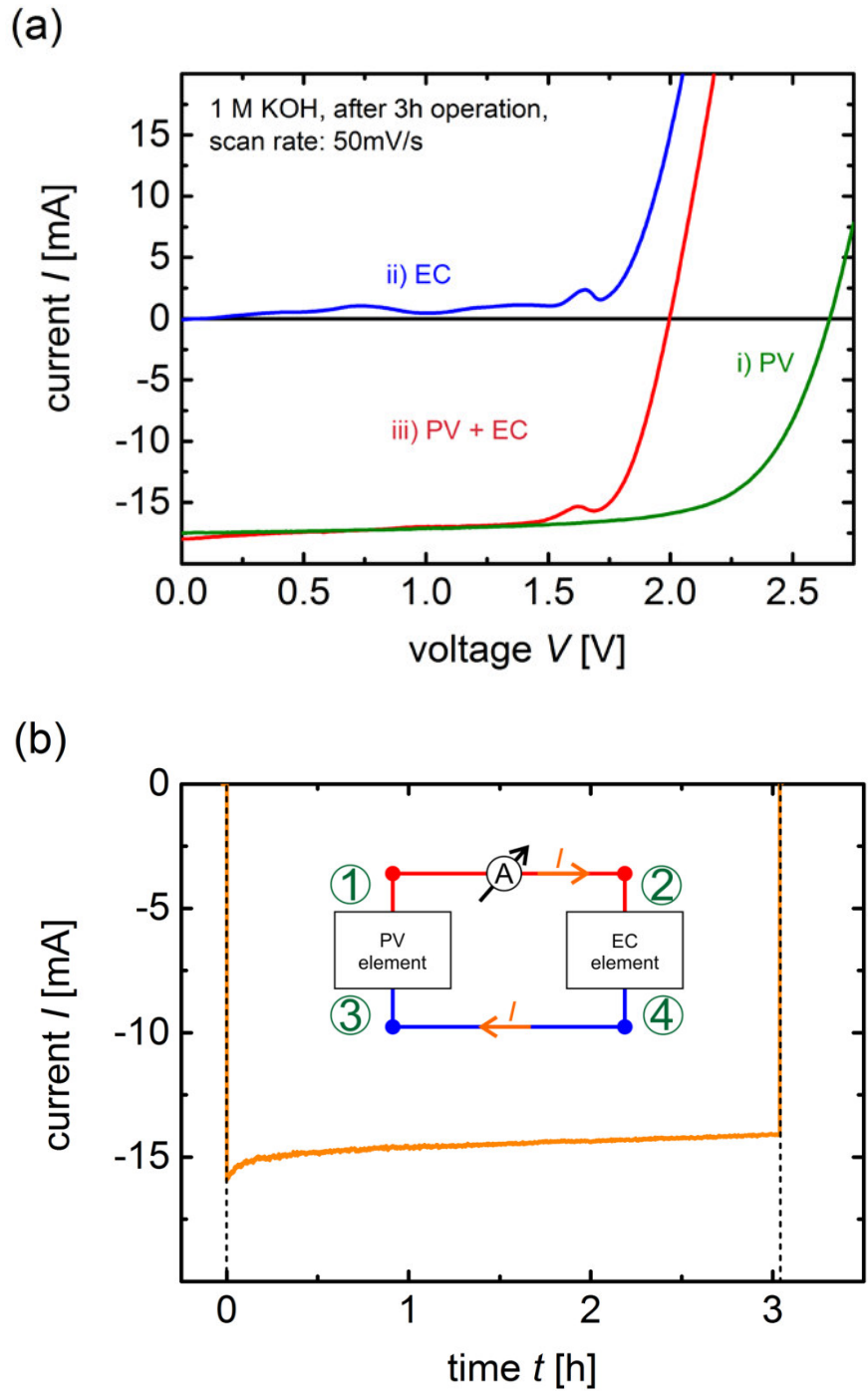

**Figure 4: Current-voltage (*I-V*) characteristics of device #1 shown in Figure 3(a) in different operating modes. (a) Individual and combined *I-V* curves of the PV and EC elements after 3 hours of operation under illumination. The *I-V* sweeps were all done starting from $V = 0$ V. (b) Operating current of device #1 under illumination as a function of time with the operating mode depicted by the inset.**

From Figure 4(a) it can be seen that in each operating mode the individual sub device characteristics can be evaluated. The superposition of the individual characteristics in operating mode i) and ii) match the *I-V* curve in the combined PV + EC operating mode iii).

The current through the system during photovoltaic water splitting is plotted in Figure 4(b) over the operating time. The measurement mode is illustrated by the inset. With Faraday´s law of electrolysis and the ideal gas law the theoretical evolution rate of oxygen and hydrogen can be calculated from the current with the following relation.

$$r = \frac{dV}{dt}\frac{1}{A} = \eta_F \frac{RT}{F \times p \times z \times A} I \qquad (4)$$

The gas rate $r$ is related to the current $I$ multiplied with a factor consisting of the ideal gas constant $R$, the temperature $T$, Faraday´s constant $F$, gas pressure $p$, and the ratio between charge transfer and gas molecules formed $z$, which is 1.33 for water splitting. For the sake of comparability the gas rate was normalized by the illuminated device area $A$. Furthermore, the factor $\eta_F$ can be regarded as the faradaic efficiency which is ideally unity.

The theoretical gas rate from the current measurement can be compared to the gas rate that was measured by collecting the co-evolved reaction products within a bell jar. For simplicity reasons a membrane was not used in this experiment and hydrogen and oxygen were collected together. Figure 5 shows both values as a function of time.

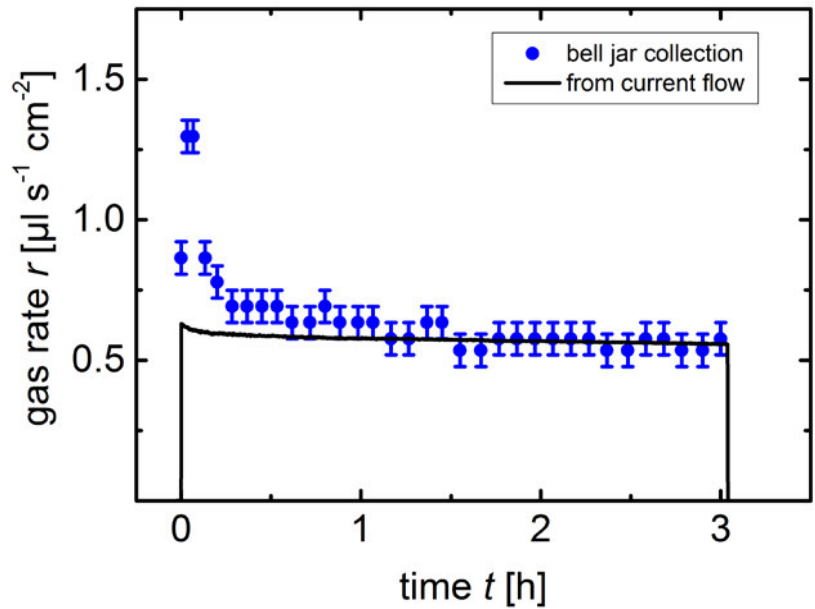

**Figure 5: Gas flow rate $r$ of device #1 as a function of time for a theoretical calculated gas rate from the current measurement (black curve) and gas rate evaluated from collection of reaction products with a bell jar (blue circles). The systematic error originates from the evaluation of volume scale bar and time stepping.**

It can be seen that the calculated and the experimental values merge after a certain time delay which is an indication that the faradaic efficiency of the device was in fact close to unity.

**Durability (device #2)**

For the investigation of the device durability a fully integrated photovoltaic water splitting device was prepared with nickel-filled epoxy as electrical connection between PV and EC element, as intended in the design in Figure 2. A photograph of this device (referred to as device #2) glued onto a glass substrate for clamping into the measurement setup is shown in Figure 6.

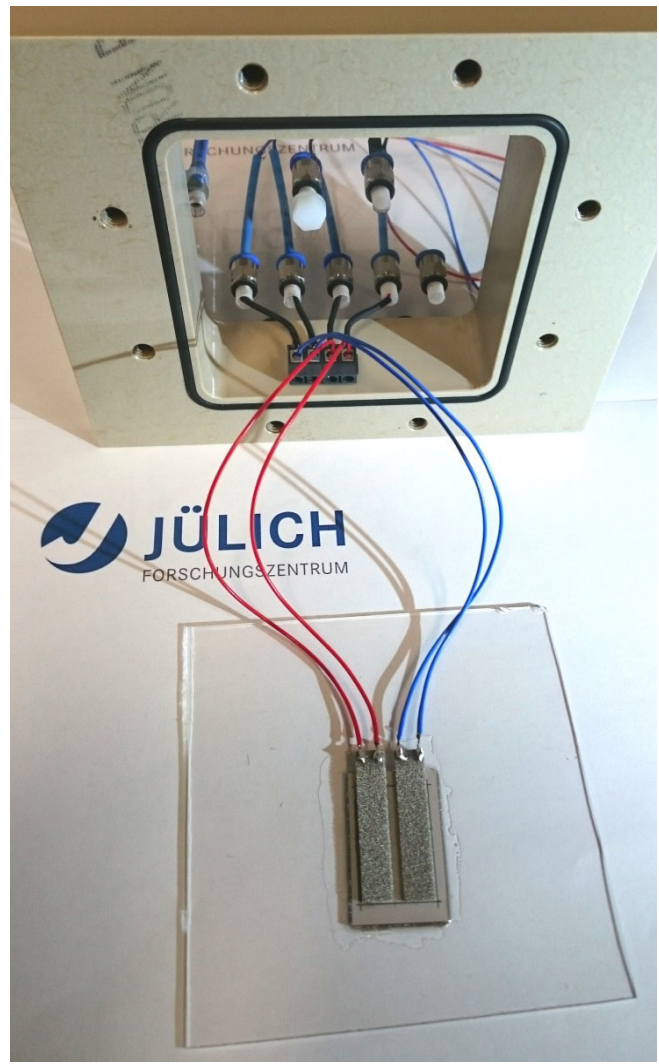

**Figure 6: Photograph of a device consisting of one base unit with three a-Si:H solar cells connected in series (device #2). Both, anode and cathode are placed on the back side and consist of nickel-foam immersed in 1M KOH. For the measurement of the electrical properties a four-point connection is soldered to the EC electrodes. A series connection of three cell stripes, each with a length of 29.8 mm and width of 5 mm lead to a device area of 5.3 cm². The device was clamped into the gas tight fixture made of polyether ether ketone.**

For the characterization of the electrical properties two wires were soldered to each EC electrode to avoid any influence of the contact resistances. Thus, only the parallel connection of both PV and EC elements can be evaluated. During photovoltaic water splitting it is only possible to monitor the operating voltage of the device. However, without the electrolyte it is possible to characterize the PV element.

Figure 7(a) shows the current-voltage (*I-V*) characteristics of the device under illumination in 1M KOH (PV+EC) as well as without the electrolyte (PV) before and after 20 hours of operation.

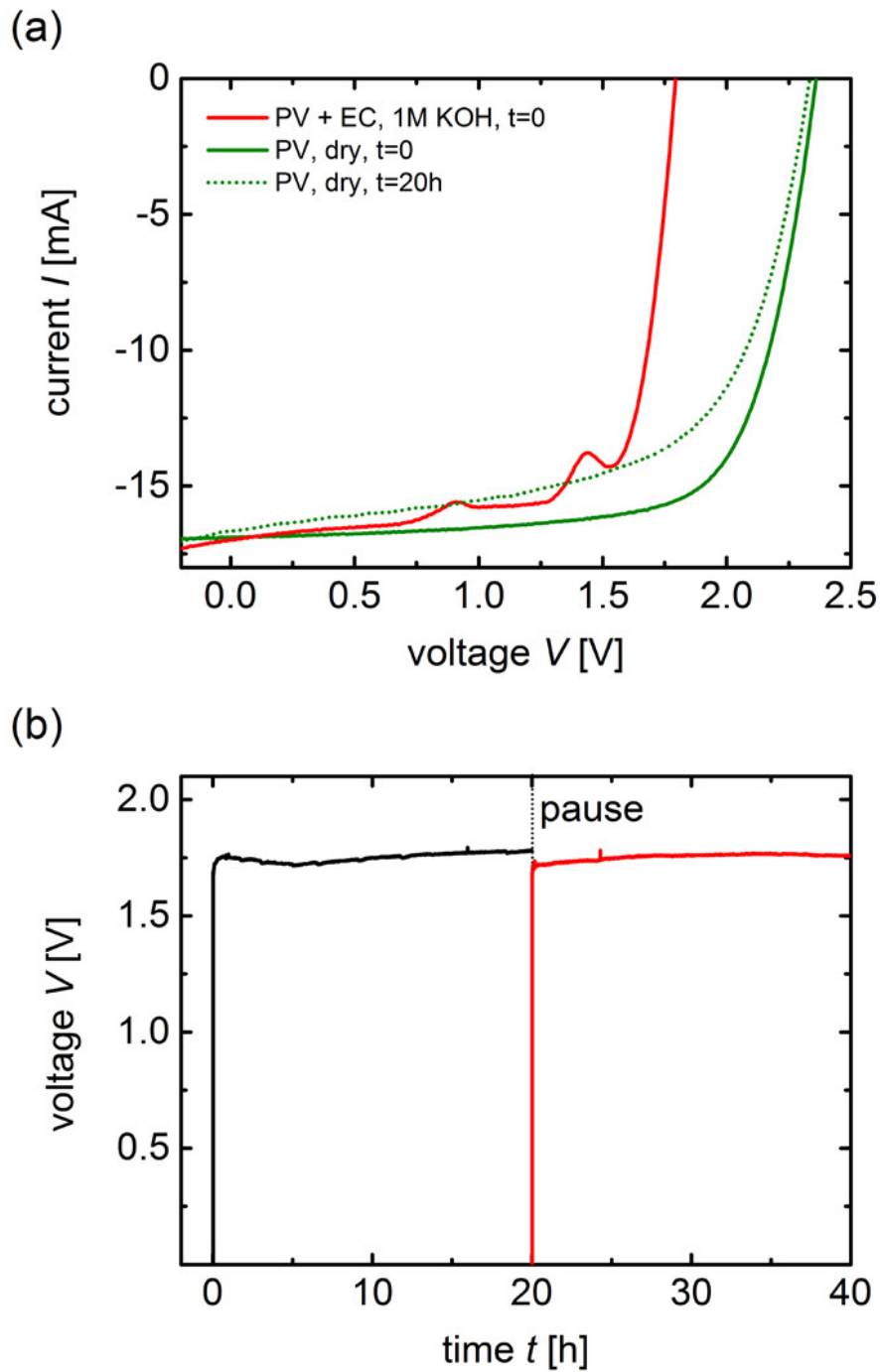

**Figure 7: Electrical measurements of device #2. (a) *I-V* characteristics of the device during illumination in 1M KOH as well as the PV elements characteristics before and after 20 hours of operation. (b) Operating voltage of the device during illumination in 1 M KOH as a function of the time.**

A degradation of the photovoltaic *I-V* characteristics was observed after 20 hours, mostly due to a reduction of the fill factor. This can be attributed to the well-known Staebler-Wronksi effect for amorphous silicon technology[43]. The degradation rate is initially stronger before it saturates for higher operation times[44]. The electrolyte was removed for the characterization after 20 hours and the experiment was resumed with a fresh 1M KOH solution.

The *I-V* curve of the whole system (PV+EC) shows an operating voltage ($I = 0$ mA) of approx. 1.75 V. The nickel-foam exhibits a very large surface to volume ratio of approx. 6900 m²/m³. The large surface area results in a low current density and, consequently, in a low overpotential. For an optimization of the device, the PV element's properties need to be adjusted to fit the EC characteristics and *vice versa*. The integrated series connection provides an additional degree of freedom for the adjustment of the *I-V* characteristics of the PV element.

In Figure 7(b) the operating voltage is monitored during water splitting over the accumulated period of 40 hours. The constant voltage indicated excellent stability of the device during this period. After 20 hours the experiment was halted for characterization and resumed afterwards for an additional 20 hours.

The co-evolved gaseous reaction products were collected. Figure 8 shows the evaluated gas rate as a function of time.

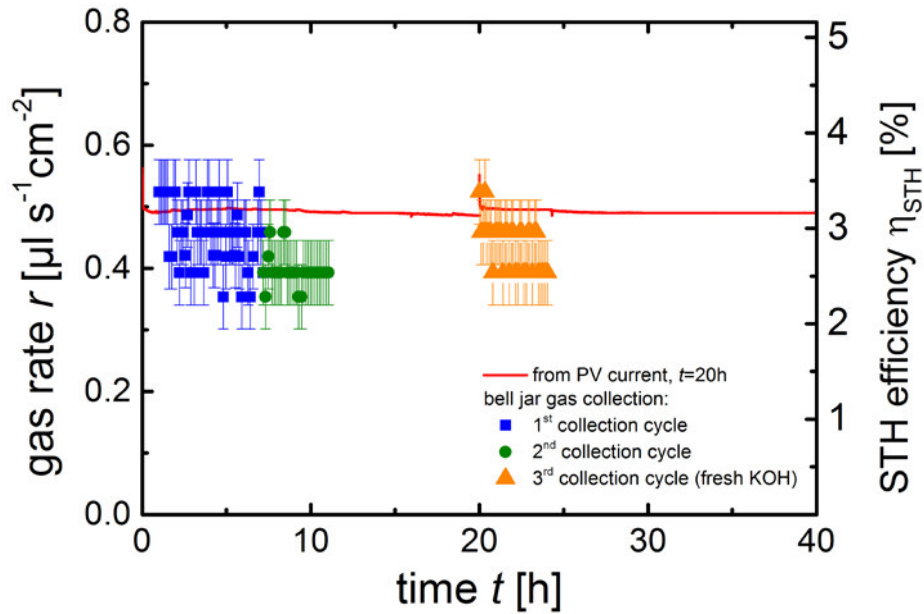

**Figure 8: Gas rate *r* evaluation of device #2 with three series connected a-Si:H solar cells and nickel-foam electrodes in 1M KOH under illumination. After 20 hours the electrolyte was removed for the characterization of the device and the experiment was resumed in a fresh electrolyte. The gas rate was evaluated from three gas collection cycles as indicated by the different symbols. The solid line indicates the**

**theoretical gas rate calculated via Equation 4 using the current extracted from the operating voltage of the PV element's *I-V* curve after 20 hours of operation (see Figure 7(a)). The STH efficiency, as calculated with Equation 5, is shown on the right hand y-axis.**

The different symbols correspond to three gas volume collection sequences. After each sequence the gas was removed from the bell jar. The evaluated gas rate was in the range of 0.35 to 0.55 µl s$^{-1}$ cm$^{-2}$ during the experiment. A certain reduction is observed as seen in the second evaluation (cf. circles in Figure 8). However, to some degree the initial gas rate was recovered after exchange of the electrolyte (cf. triangles in Figure 8).

The current through the system cannot be extracted due to the structure of the device. However, with the characterization of the *I-V* curve of the PV element it is possible to correlate the operating voltage to the equivalent operating current. The line curve in Figure 8 indicates the theoretical gas rate calculated from this current via Equation 4. For the complete time period of 40 hours the PV element´s *I-V* curve of the device after 20 hours was used for current operating point extraction.

It can be seen that the calculated gas rate correlates well to the experimental gas rate evaluation. However, there is a slight overestimation of the calculated gas rate. This discrepancy may be explained by a time-dependent change of the PV element's *I-V* curve. For instance, the system temperature differs before, during, and after device operation. For a reliable assessment of the device performance a quantification of the evolved gases is favorable.

The gas rate of the system can be used to evaluate the solar-to-hydrogen (STH) efficiency of the device.

$$\eta_{STH} = \frac{2}{3} \frac{r \times \Delta G}{P_{IN} \times V_m} \tag{5}$$

In Equation 5, $r$ is the gas rate, the factor 2/3 corresponds to the hydrogen to oxygen ratio, $\Delta G = 237$ kJ/mol is the change in Gibb's free energy per mole of H$_2$, $V_m = 24.8$ l/mol is the molar volume of an ideal gas, and $P_{IN} = 100$ mW/cm$^2$ is the illumination power density. From Equation 5 follows that a STH efficiency of 1% corresponds to a gas rate of $r = 0.155$ µl s$^{-1}$ cm$^{-2}$.

Furthermore, the STH efficiency can be calculated from the current *I* through the system[45].

$$\eta_{STH} = \frac{1.23V \times I \times \eta_F}{P_{IN} \times A} \tag{6}$$

In Equation 6, $\eta_F$ is the faradaic efficiency. The knowledge of the spectrum and the intensity of the used solar simulator is crucial for a correct determination of the STH efficiency[46].

From Figure 8 a mean value of the device area STH efficiency of approx. 3% was observed for device #2 for more than 20 hours. This corresponds to an active area STH efficiency of 3.3% (see Figure S13 in the supplements). A better matching of the PV and the EC element's characteristics could increase the STH efficiency. Furthermore, the electrochemical activity of the nickel-foam can be increased significantly by a modification with suitable catalysts[47–49]. The utilized a-Si:H single junction solar cells could ideally yield a STH efficiency of 5.4% (calculated from data in Figure S9).

To estimate the potential of our concept, that enables more possibilities with regard to the *I-V* output characteristics, a simple calculation was made for various PV technologies with the best reported catalysts. Table 1 shows the theoretical solar-to-hydrogen efficiencies achievable with reported values from the literature.

**Table 1. Calculated maximal solar-to-hydrogen efficiencies using state-of-the-art PV technology. For the EC element NiMo and NiCo were used as catalysts with their respective overpotentials at 10 mA/cm² from McCrory et al.[8]. Any *IR*-drop in the electrolyte and additional losses were neglected which lead to an overall water splitting voltage of 1.7 V. The current density at 1.7 V was divided by the number of cell stripes.**

| PV Technology | Current density divided by number of cell stripes | Number of solar cell stripes | Potential Solar-to-hydrogen efficiency | Reference |
|---|---|---|---|---|
| Silicon triple junction | 7.7 mA/cm² | 1 | 9.5% | Urbain et al.[14] |
| Perovskite solar cell | 11.7 mA/cm² | 2 | 14.4% | Yang et al.[42] |
| Cadmium-Telluride (CdTe) | 10.1 mA/cm² | 3 | 12.4% | Green et al.[50] |
| Cu(In,Ga)Se$_2$ (CIGS) | 11.5 mA/cm² | 3 | 14.2% | Green et al.[50] |

Depending on the PV technology the number of series connected cells was varied to achieve matching to the EC element's required voltage of 1.7 V (NiMo and NiCo catalysts and no additional losses). The output voltage of the solar cell was multiplied by the number of cell stripes while the current density was divided accordingly. Interconnection losses were neglected.

It should be emphasized that most of the shown PV technologies cannot sustain the required voltage of 1.7 V without a series connection. With our design concept, large area, scalable, efficient photovoltaic water splitting devices are feasible for all thin-film PV technologies where a series connection can be realized, e.g. by laser processing[51–54].

**Large area device (device #3)**

In the previous section the successful realization of one base unit of our concept was demonstrated and showed stable operation over 40 hours. Next, the upscaling of the concept with multiple adjoining base units will be presented (device #3). In contrast to the previous experiments (device #1 and #2), each base unit consisted of two series connected a-Si:H/μc-Si:H tandem junction solar cells. This allows for closer packing of the base units on the substrate. In total 13 adjoining base units were prepared on a $10 \times 10$ cm² substrate. The overall area of the device was $A = 64$ cm². Figure 9 shows a photograph of this device after preparation.

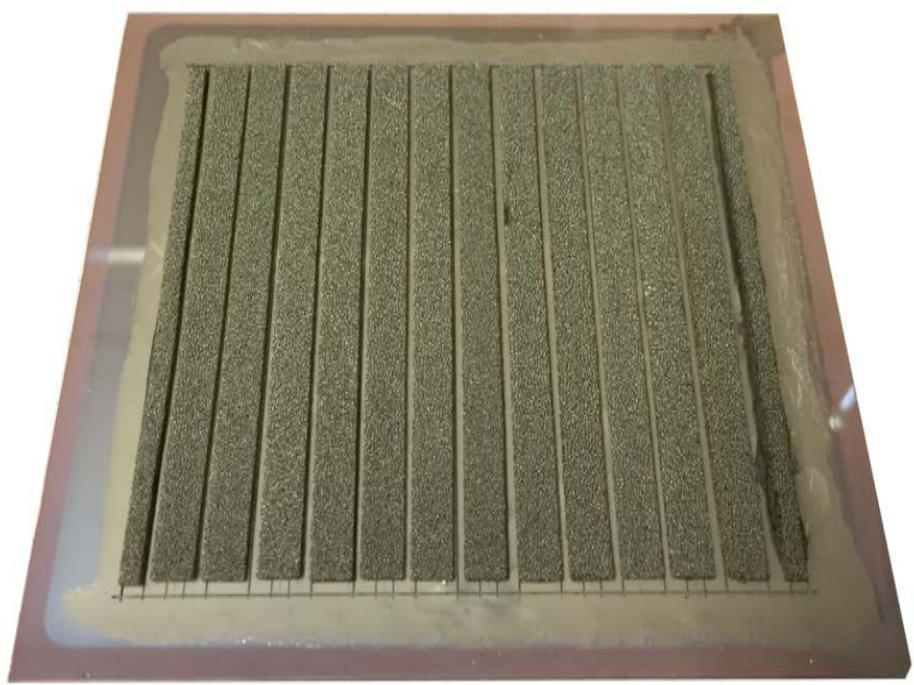

**Figure 9: Photograph of a large scale photovoltaic water splitting device (device #3). The total device area was 64 cm² with an active area of 52.8 cm². Each base unit consists of two series connected a-Si:H/μc-Si:H tandem solar cells with a cell stripe width and length of 2.5 mm and 80 mm, respectively. Thirteen base units were neighboring on a $10 \times 10$ cm² substrate. The back-end was made of laser-cut nickel-foam elements for both, cathodes and anodes.**

The cell stripe width for each base unit was 2.5mm which is not ideal in terms of solar module efficiency[55] but rather chosen to illustrate the successful upscaling of the concept by having a multitude of adjoining base units. The cell stripe length was 80 mm and the interconnection width was 300 μm. Since the insulating epoxy was deposited manually a rather wide fillet to the front contact of 2 mm was required which decreases the active PV area. However, with an optimized laser patterning process a fillet width as narrow as 100 μm is possible which increases the active area[56,57].

For this fully integrated device #3, monitoring of the voltages is not feasible and due to the symmetric design of the interconnection the base unit´s PV element properties are not accessible. Therefore, proper upscaling behavior was evaluated by the measurement of the gas rate. Figure 10 shows the overall collected gas volume as well as the gas rate as a function of the operating time.

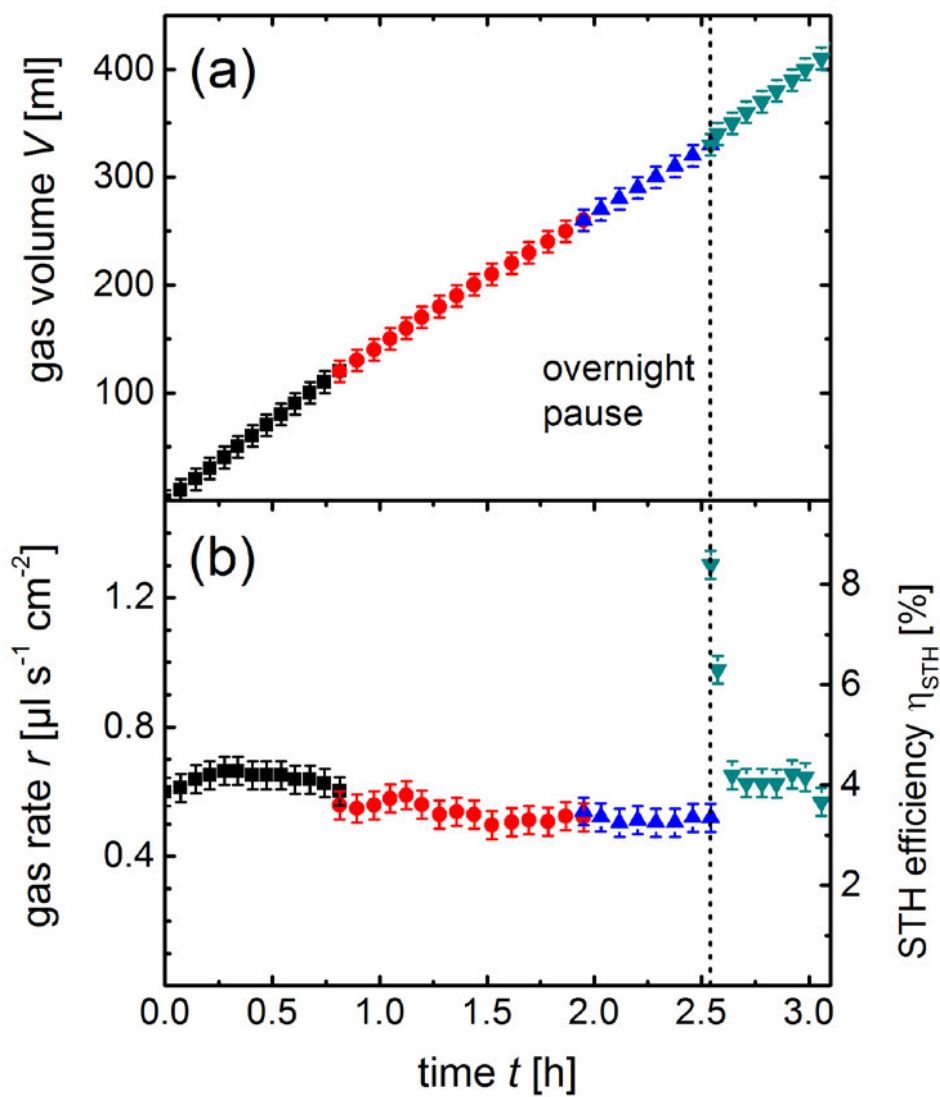

**Figure 10: Collected gas volume (a) and gas rate/STH efficiency (b) of an upscaled water splitting device (device #3) as a function of time. The total device area was 64 cm² with thirteen neighboring base units. Each base unit consisted of two series connected a-Si:H/µc-Si:H tandem solar cells. The cell stripe width was 2.5 mm with a length of 80 mm. The individual symbols/colors in the graphs correspond to gas volume collection sequences after which the collection bell jar needed to be emptied. The dashed line marks where the experiment was paused overnight.**

Due to the high overall gas rate the bell jar needed frequent clearing to reset the collection as indicated by the different symbols in Figure 10. A decrease of the gas rate over time was observed. However, after a pause overnight (marked with a dashed line in Figure 10) of the experiment the initially higher gas rate was partially recovered which indicates a variation of the operating condition over time rather than a degradation of the device.

The successful upscaling of the concept becomes apparent when the gas rates of device #2 ($r \approx 0.4$ µl s$^{-1}$cm$^{-2}$, see Figure 8) and device #3 ($r \approx 0.6$ µl s$^{-1}$cm$^{-2}$, see Figure 10) are compared. The device area was increased from 5.3 cm$^2$ to 64 cm$^2$ while the gas rate was maintained. The slightly higher gas rate of the upscaled module can be attributed to the better matching of the *I-V* characteristics of PV and EC element in the case of the series connection of two tandem solar cells. Such base unit usually exhibits a higher current density compared to three series connected a-Si:H single junction solar cells. The extracted device area STH efficiency for device #3 was approx. 3.9%. Considering only the active area a STH efficiency of 4.7% was achieved. This is remarkable for non-optimized device geometries in terms of cell stripe width and catalyst dimensions. The utilized a-Si:H/µc-Si:H tandem solar cells could ideally yield a STH efficiency of 6.6% (calculated from the data in Figure S10) assuming an operation voltage of 1.7 V.

**Summary and outlook**

In this study we introduced and discussed concepts for the upscaling of artificial photosynthesis from laboratory scale to large areas. All concepts feature continuous mirroring and/or repetition of a single base unit to achieve the scalability.

The realization of one concept was showcased using thin-film photovoltaic technology and non-precious metal catalysts in alkaline solution. An integrated series connection by means of laser material processing was used. Thus, the concept allows for the application of any thin-film photovoltaic technology where the voltage of a single junction would not be sufficient to sustain electrolysis.

The proof of concept was presented first, with a single base unit that was modified to access all relevant photovoltaic and electrochemical properties individually. Afterwards, a second single base unit was created which exhibited excellent stability for more than 40 hours of operation under illumination in 1M KOH. Finally, an upscaled device with an area of 64cm² and multiple base units has proven the successful scalability of the concept by a similar gas rate per unit area between the upscaled and single base unit device. To our knowledge, this is the first reported scalable and wireless photovoltaic water splitting device on this scale. For the upscaled device a solar-to-hydrogen efficiency of approx. 3.9% was measured.

We believe that these results may encourage the successful application of photo-electrochemical water splitting to system scales that are relevant for a combined renewable energy generation and storage which is gaining substantial importance in the future.

Beyond the generation of hydrogen artificial photosynthesis also addresses the reduction of $CO_2$[58] or the creation of a sustainable closed carbon cycle (e.g. solar fuels). In this broader context many chemical challenges are involved[4]. One key challenge is the supply of a sufficient voltage to sustain the electrochemical reactions. This aspect is successfully addressed by the proposed concepts.

## Methods

### Solar cell deposition

For all experiments in this work commercially available 1.1 mm thick glass substrates coated with fluorine-doped tin dioxide ($SnO_2$:F) as transparent conductive oxide (TCO) from the Asahi Glass Company (type U) were used as front contact material[59].

We applied state-of-the-art thin-film silicon technology for the deposition of p-i-n single junction solar cells from hydrogenated amorphous silicon (a-Si:H) and multi-junction p-i-n/p-i-n solar cells from hydrogenated amorphous and microcrystalline silicon (a-Si:H/µc-Si:H) by Plasma Enhanced Chemical Vapor Deposition (PECVD)[60].

A layer stack consisting of aluminum-doped zinc oxide (ZnO:Al) and silver (Ag) was deposited as back contact material by RF magnetron sputtering[61].

### Laser processing

For the integrated series connection[22] Nd:YVO$_4$ Q-switched diode-pumped solid state laser sources from ROFIN (type PowerLine E) were used which exhibit a pulse duration between 7-12 ns (FWHM). For the front contact insulation process (P1) the third harmonic with a wavelength of 355 nm was used. A pulse repetition frequency of 15 kHz was applied and the average power of the laser was 210 mW. Focusing was done with an f-theta lens (f = 108 mm) which lead to a beam spot radius of 19 µm. A laser source with the second harmonic and a wavelength of 532 nm was used for the removal of the absorber (P2) as well as for the back contact removal process (P3). The pulse frequencies were 17 kHz and 11.5 kHz, respectively. For P2 the average power was 430 mW while 290 mW was used for P3. For both processes a lens system with a focal length of f = 300 mm was used to focus the laser beam which led to a beam spot radius of 60 µm. The three processes P1-P3 required an overall interconnection width of 300 µm. All processing was done through the glass side. The additional processes required for the short-circuiting between front and back contact were modified P2 and P3 processes. Details on laser processing can be found in the supplemental data.

**Device Assembly**

For corrosion protection and electrical insulation of the system´s back-end a chemically resistant epoxy from Loctite (type 9483) has proven to be a suitable two-component adhesive for manual patterning of the PV element´s back side. The contact to the PV element was masked by adhesive tape which was removed after homogenous application of the epoxy and before annealing on a hot plate for 20 mins at a temperature of 100°C.

Subsequently, a one-component nickel-filled epoxy from Alfa Adhesives (type E10-102) was applied onto the exposed contacts of the PV element. On top of this adhesive conductive epoxy, 1.4 mm thick nickel-foam electrodes (RECEMAT BV, Ni-5763) were attached and the whole device was annealed for 120 mins at 125°C. The individual nickel-foam stripes were laser-cut to the required width and length by a $CO_2$ laser tool.

**Characterization**

Setup

Single base units were glued with epoxy (Loctite 9483) onto a 10 × 10 cm² glass substrate and clamped into a hermetically sealed sample holder made from polyether ether ketone (PEEK). Four wires were fed into the chamber as well as tubing for electrolyte filling and extraction of the gaseous reaction products during the water splitting experiments. For the upscaled water splitting devices on 10 × 10 cm² substrates an additional glass substrate for mechanical support was not required.

Single base units were illuminated with a small area class ABB solar simulator from Newport (type LCS-100 model 94011A). The distance between light source and device was adjusted to generate an incident illumination intensity of approx. 1kW/cm² with the help of reference measurements using a spectrophotometer together with calibrated solar cells. The spectrum of the sun simulator was measured simultaneously with the reference. For large scale devices an in-house built large area sun simulator was used. All spectra can be found in Figure S5. Additionally, various photographs of both measurement setups are shown in Figure S6-S8.

Electrochemical characterization

All experiments were conducted in an aqueous 1M potassium hydroxide (KOH) solution prepared from analytical grade chemicals (Merck). The deionized water (Millipore) used for the preparation of the solution was scrubbed with air for 15 mins for the saturation with oxygen. To illustrate the influence of the concentration on the system a comparison of the EC element's *I-V* characteristics in 0.1M and 1M KOH is shown in Figure S3.

A potentiostat/galvanostat from Gamry Instruments (Reference 600+) was used for the electrochemical characterization as well as for voltage monitoring over time and *I-V* characteristics measurements of the PV element. The potentiostat/galvanostat was either used in a two-wire or four-wire sense mode. If not specifically stated otherwise, a scan rate of 50 mV/s was used for all measurements with an open-circuit delay of 30 s.

For the monitoring of the time-resolved current of device #1 during illuminated operation a multimeter from Keithley (model 2000) was used.

<u>Measurement of the co-evolved gases</u>

Hydrogen and oxygen were collected with a bell jar or inverted burette with two different maximum volumes of 100 ml and 1000 ml. A camera was focused onto the bell jar and took time-lapse photography so that the generated gas volume could be evaluated visually together with the timestamp for prolonged operation durations. Each bell jar exhibits a scale stepping that was 1ml and 10ml, respectively. We estimated the reading error to be ±1 ml and ±2 ml. Before the measurement the deionized water that is used as a gas barrier in the gas collection system was scrubbed with air for 15 mins. Figure S7 shows a photograph of the gas collection apparatus.

## Acknowledgements

The authors would like to thank H. Siekmann, A. Bauer, G. Schöpe, J. Kirchhoff, and C. Zahren for their contributions to this work. The research was partly financially supported by the Deutsche Forschungsgemeinschaft (DFG) Priority Program 1613 (SPP 1613), and by the German Bundesministerium für Bildung und Forschung (BMBF) in the network project Sustainable Hydrogen (FKZ 03X3581B).

# SUPPLEMENTARY INFORMATION

# From leaf to tree: upscaling of artificial photosynthesis

*Bugra Turan\*, Jan-Philipp Becker, Félix Urbain, Friedhelm Finger, Uwe Rau, and Stefan Haas*

IEK5-Photovoltaik, Forschungszentrum Jülich GmbH, 52425 Jülich, Germany
E-mail: b.turan@fz-juelich.de

## 1. Supplemental *I-V* characteristics data for: Device #1

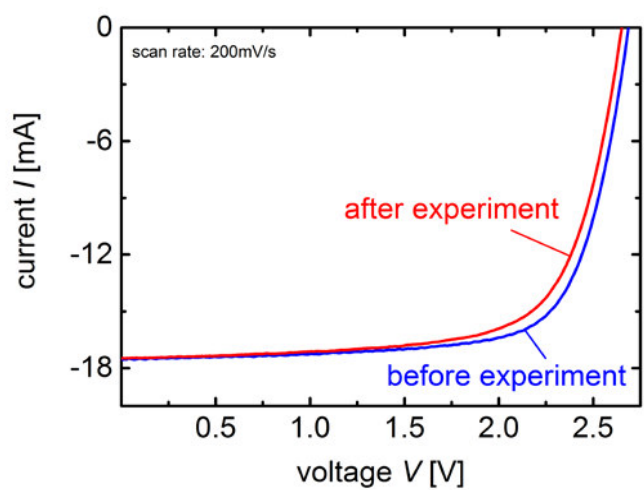

**Figure S1.** Plot of the PV element´s *I-V* characteristics of device #1 before and after 3 hours of operation under illumination. A reduction of the fill-factor and a lowered open-circuit voltage can be seen from the curve after operation. Device degradation as well as increased temperatures could be responsible for such behavior since there was no control over the device temperature. Both measurements start from $V = 0$ V and a scan rate of 200 mV/s was employed.

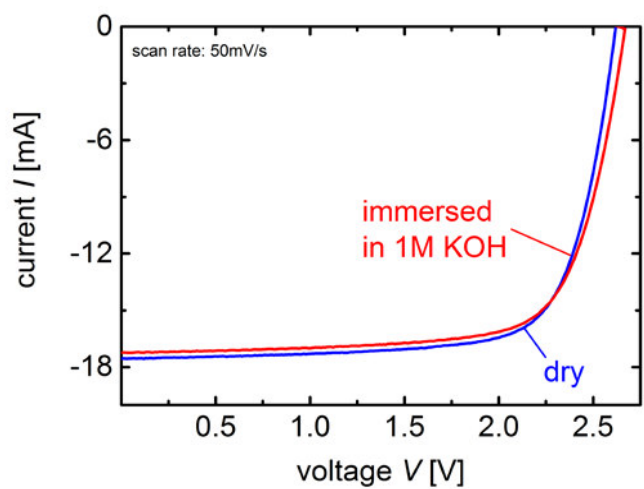

**Figure S2.** Plot of the PV element's *I-V* characteristics of device# 1 dry and immersed in 1M KOH before operation. Both curves have differences in short-circuit current as well as in the open-circuit voltage. One explanation for these differences can be identified by a change of the device temperature by cooling due to the electrolyte which acts as a heat sink to a certain degree. Both measurements start from $V = 0$ V and a scan rate of 50 mV/s was employed.

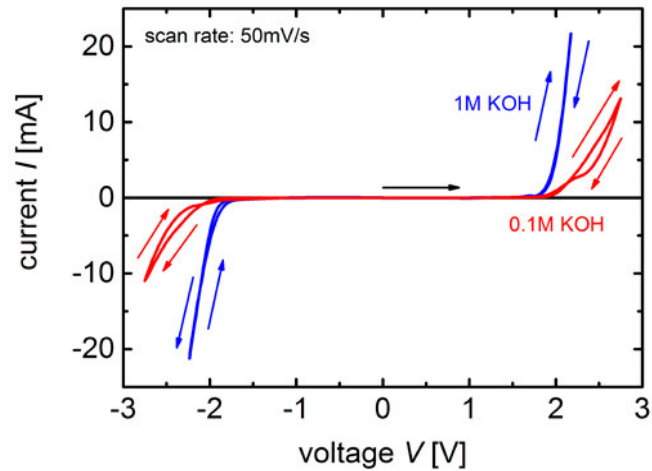

**Figure S3.** Influence of the KOH concentration on the EC element´s *I-V* characteristics of device #1. The arrows indicate the scan direction. Both measurements start with *V* = 0 V into positive voltages (see black arrow). The effect of the higher ionic conductivity of the measurement in 1M KOH solution is clearly visible from the much steeper slope above 1.8 V. A scan rate of 50 mV/s was employed.

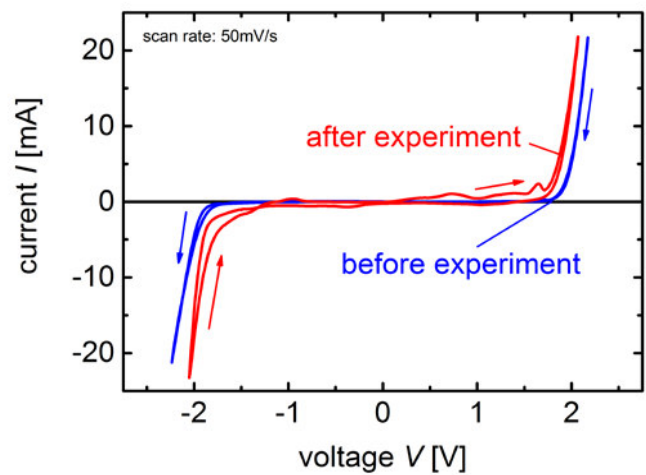

**Figure S4.** Comparison of the EC element´s *I-V* properties from device #1 before and after 3 hours of operation under illumination in 1M KOH. After operation a lower overpotential is observed which can be explained be chemical changes of the nickel-foam surface. Oxidation and reduction potentials can be identified that were initially not apparent. A scan rate of 50 mV/s was employed. Please note that no preconditioning of the nickel-foam was done prior to the experiment.

## 2. Measurement Setup:

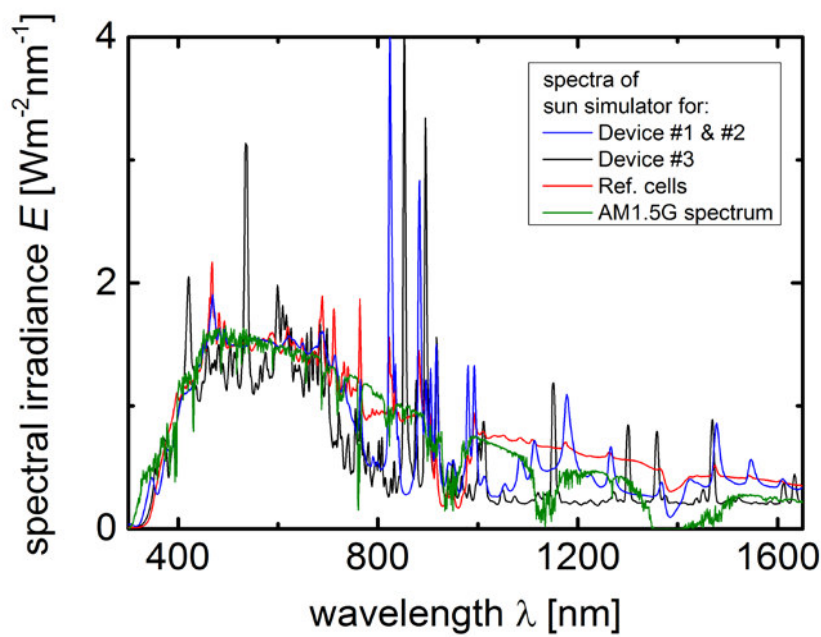

**Figure S5.** Plot of spectral power density vs. wavelength of different sun simulators that were used for water splitting experiments and reference solar cell measurements. All illumination sources were steady state lamps. Shown in green is the AM1.5G reference spectrum. There is a good agreement observed between the reference and the sun simulators for the relevant wavelength region of thin-film silicon (400 nm to 1100 nm).

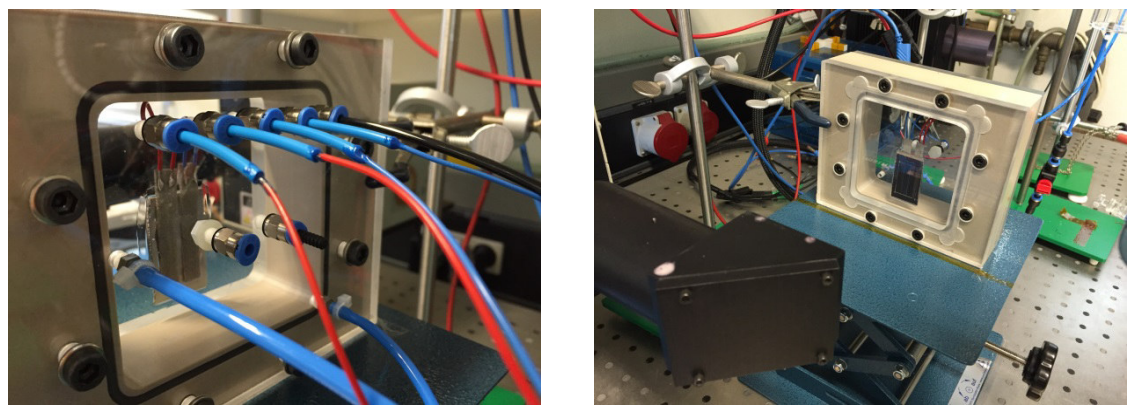

**Figure S6.** (left) Photograph of the measurement setup with the holder made from polyether ether ketone (PEEK) and the single base unit called device #2. Wires were soldered to the anode and cathode for four-wire sensing and fed-through with a gas-tight sealing (blue connectors). Co-evolved reaction products were extracted by the tube in black. The two thicker tubes in blue were used for filling and draining of the electrolyte. (right) Photograph of the setup from the front side with the sun simulator (cf. blue curve in Figure S5). The device under test was glued with a transparent epoxy onto a glass substrate for an easy, gas-tight, fixture onto the holder.

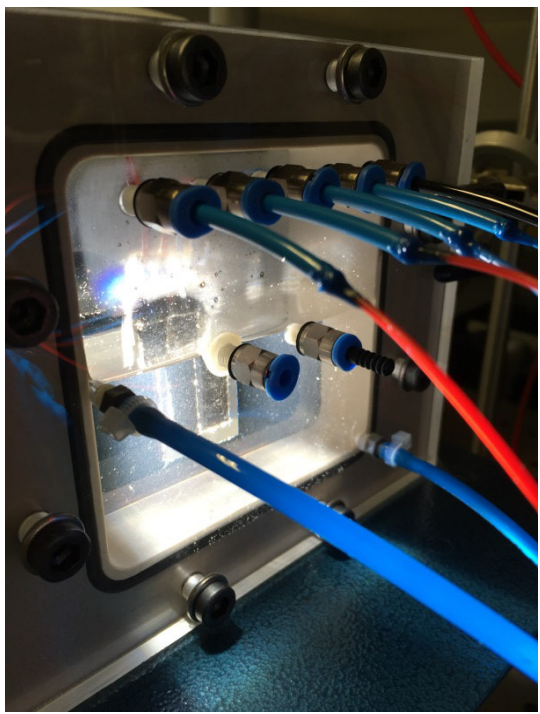 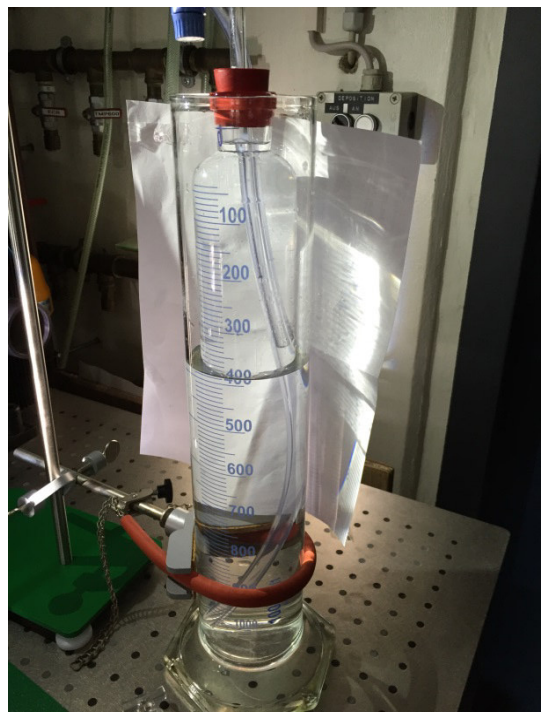

**Figure S7.** (left) Measurement setup with the single base unit device #2 fixed into the sample holder under illumination from the front side. The fixture was filled with 1M KOH as electrolyte. (right) Photograph of a 1000 ml bell jar (or inverted burette) for collection of the co-evolved gas products (see black tube in the left picture). The gas rate was evaluated by time-lapse photography of the fluid level change over time.

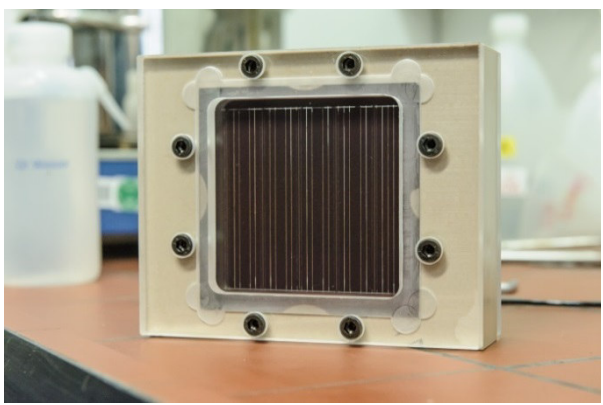 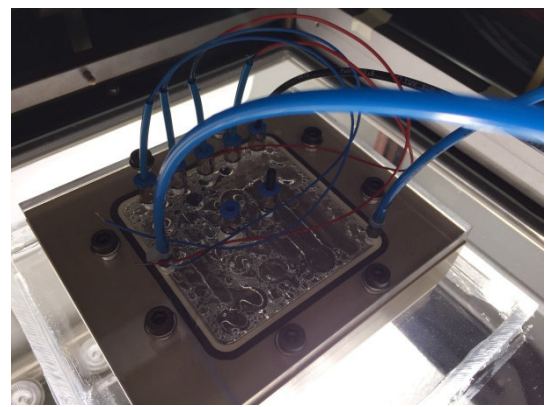

**Figure S8.** (left) Photograph of the sample holder shown from the front side assembled with the upscaled 10 cm × 10 cm module (8 cm × 8 cm aperture area) similar to device #3. (right) Photograph of the measurement setup for the upscaled water splitting devices with an in-house built large area sun simulator. The device was illuminated from the bottom. Voltage monitoring was not applicable for this device.

## 3. Reference solar cells and equivalent circuits for the devices #2 and #3:

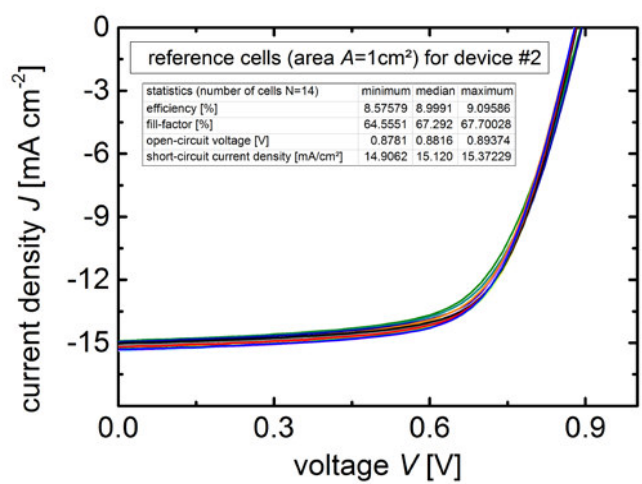

**Figure S9.** Solar module *J-V* properties of reference a-Si:H solar cells (area $A = 1$ cm²) that were deposited in the same deposition sequence as the stability test base unit named device #2. A total of 14 solar cells were measured on a 10 cm × 10 cm substrate. The statistics in the inset table show only minor variations of the solar cell properties.

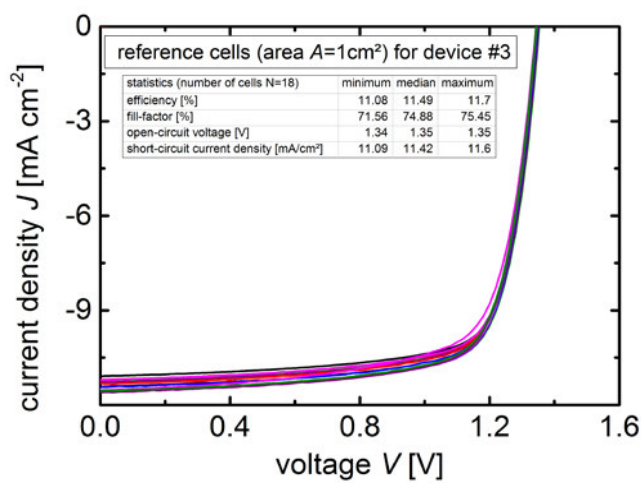

**Figure S10.** Solar cell *J-V* properties of single $A = 1$ cm² reference cells that were deposited in the same run as the upscaled water splitting device #3. A total of 18 solar cells were measured on a 10 cm × 10 cm substrate. The statistics in the inset table show only minor variations of the solar cell properties.

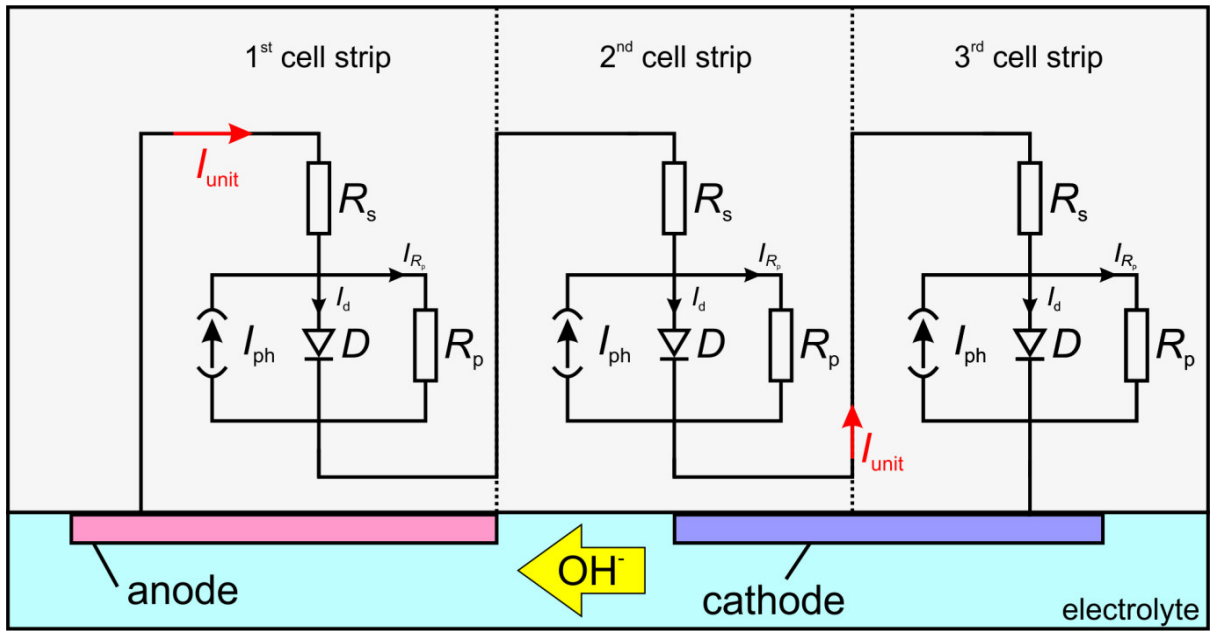

**Figure S11.** Equivalent circuit diagram of the singular base unit device #2 consisting of three single junction thin-film silicon a-Si:H solar cells connected in series. The current through the unit is depicted by $I_{unit}$. The current loop is closed by the ion flow through the electrolyte.

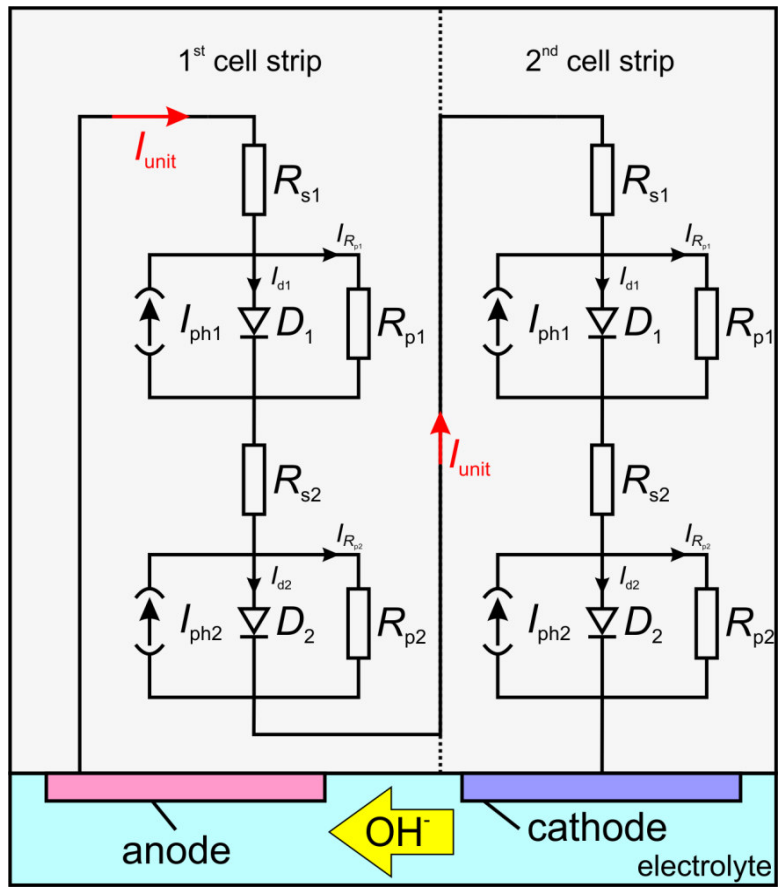

**Figure S12.** Equivalent circuit diagram of a singular base unit as in device #3 consisting of two multi-junction thin-film silicon a-Si:H/µc-Si:H solar cells connected in series. The individual sub cells are depicted by the indices 1 and 2 for the top- and bottom-cell, respectively. The current through the unit is depicted by $I_{unit}$. The current loop is closed by the ion flow through the electrolyte.

4. Miscellaneous:

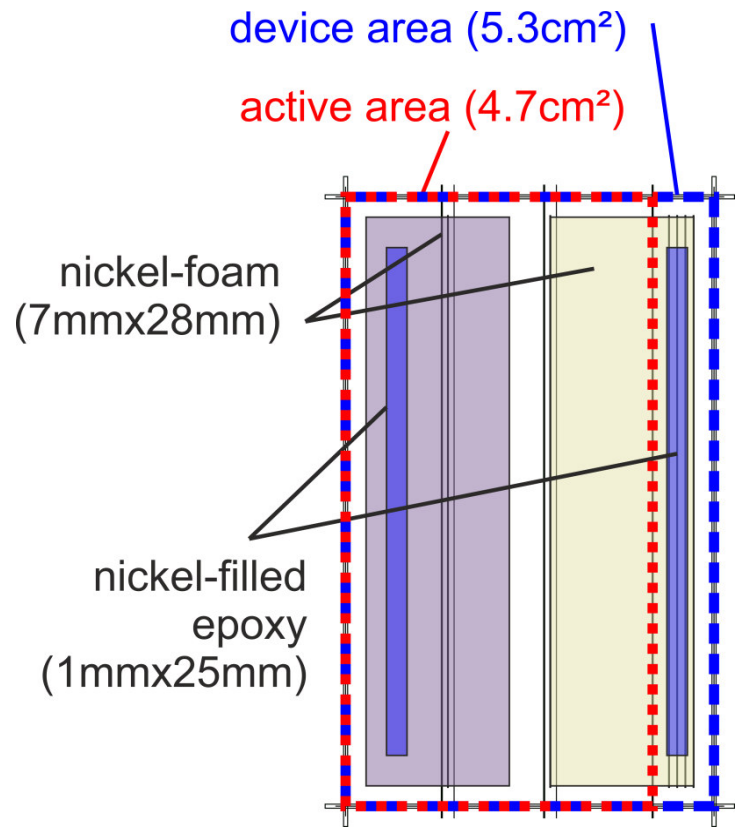

Figure S13. Schematic sketch of the laser design file for the patterning process of device #2. A transparent overlay indicates the position and size of the EC element´s anode and cathode on the back side. The areas depicted by the red and blue dashed rectangles define the active and device area, respectively. This distinction is important since the difference between both areas, due to the 2 mm wide fillet to the front contact, was required for practical reasons during device manufacture. The active area losses due to the interconnection region were neglected.

5. Laser processing:

Almost every thin-film photovoltaic technology makes use of laser processing for the series connection of solar cells. The series connection is required to lower the ohmic losses in the contacts. We often refer to it as the so called integrated series connection because the required laser processing steps are integrated in between the solar cell layer deposition steps. A first laser process is required after the front contact deposition for cell stripe definition (called P1). After the silicon deposition a second laser processing step is required to selectively remove the absorber, exposing the underlying front contact (P2). Finally, in a last process step the back contact is locally removed by the laser for the final definition of the cell stripes (P3). Figure S14 shows an illustration of a single solar module cell stripe with the interconnection structure in cross section.

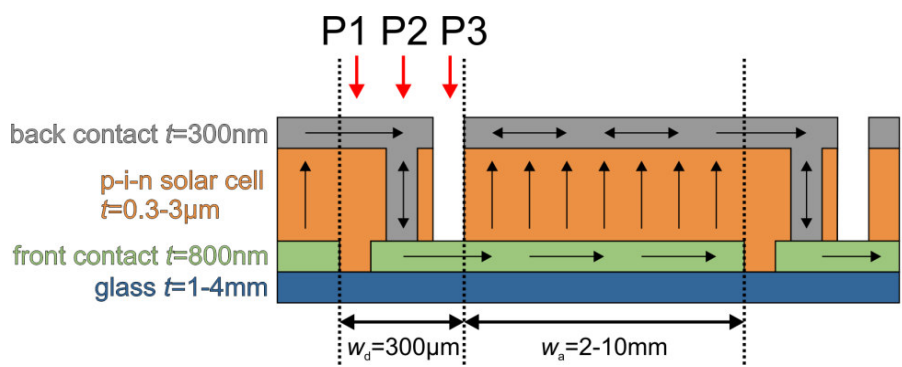

**Figure S14. Cross section illustration of a series connected thin-film solar cell in superstrate configuration. The arrows indicate the current flow direction through the solar module. The spatial width required for interconnection structure is defined by $w_d$ while the active solar cell stripe is specified by the width $w_a$.**

The individual layers are removed selectively by laser irradiation through the glass substrate leading to an ablation of the material in scribe lines, parallel across the whole substrate. For the P1 process we used a Q-switched DPSS Nd:YVO$_4$ with a wavelength of 355 nm (third harmonic) and a pulse duration between 7-10 ns. The other two ablation processes P2 and P3 were realized with the use of a similar laser source but with a wavelength of 532 nm (secondary harmonic). This wavelength is favorable since silicon is highly absorbing in this spectral range while the front contact is highly transparent ensuring selective removal without severe damages. The area required for the series connection is no longer active for charge carrier generation and for thin-film silicon typically an area of 3-5% is lost (ratio $f_d = w_d/(w_a + w_d)$). The additional laser processes are required for the fabrication of the presented device concept used similar patterning parameters as the P2 and P3 processes. For practical reasons the fillet to the front contact was 2 mm wide since manual patterning of the insulation epoxy was required (cf. Figure S13).

## 6. Time-lapse photography:

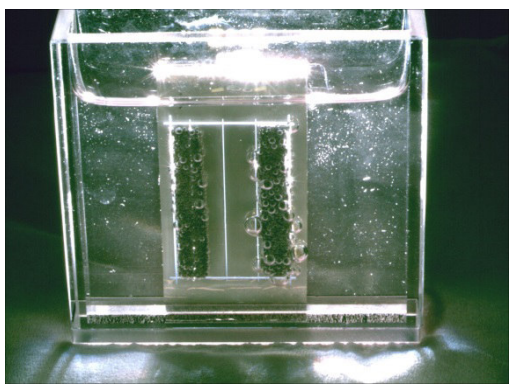

**Video 1. Time-lapse video of a single base unit device that consisted of three a-Si:H solar cells connected in series and nickel-foam stripes as the anode and cathode of the EC element immersed in 1M KOH under illumination. The total duration of the measurement outtake was 9500 s. Each photo was taken with a time interval of 10 s and a duration of 100 ms per photo is shown leading to a total video time of 95 s.**